\documentclass[12pt]{iopart}

\usepackage{iopams}
\usepackage{graphicx}

\usepackage{eucal}
\usepackage[dvips]{epsfig}
\usepackage{amssymb}

\newcommand{\be}{\begin{eqnarray}}
\newcommand{\ee}{\end{eqnarray}}
\newcommand{\bse}{\begin{subequations}}
\newcommand{\ese}{\end{subequations}}

\newcommand{\bnum}{\begin{enumerate}}
\newcommand{\enum}{\end{enumerate}}

\newcommand{\bit}{\begin{itemize}}
\newcommand{\eit}{\end{itemize}}

\newcommand{\bc}{\begin{cases}}
\newcommand{\ec}{\end{cases}}

\newcommand{\bpm}{\begin{pmatrix}}
\newcommand{\epm}{\end{pmatrix}}

\newcommand{\bvm}{\begin{vmatrix}}
\newcommand{\evm}{\end{vmatrix}}

\renewcommand{\bs}{\boldsymbol}

\newcommand{\mcal}{\mathcal}

\newcommand{\mrm}{\mathrm}

\newcommand{\ga}{\alpha}
\newcommand{\gb}{\beta}
\newcommand{\gc}{\gamma}
\newcommand{\gd}{\delta}
\newcommand{\eps}{\epsilon}%\ge schon vergeben

\newcommand{\gl}{\lambda}

\newcommand{\gt}{\theta}

\newcommand{\p}{\partial}
\newcommand{\f}{\frac}
\newcommand{\diff}{\mrm{d}}

\newcommand{\R}{\mathbb{R}}
\newcommand{\PP}{\mathbb{P}}

\newcommand{\fa}{\forall\;}

\newcommand{\kB}{k_\mrm{B}}

\newcommand{\V}{\mcal V}
\newcommand{\X}{\mcal X}

\newcommand{\G}{\mcal G}

\newcommand{\RV}{\mcal{RV}}
\newcommand{\RP}{\mcal{RP}}

\newcommand{\atanh}{\mrm{arctanh\,}}
\begin{document}

\title[Relative entropy, Haar measures and relativistic canonical velocity distributions]{Relative entropy, Haar measures and relativistic canonical velocity distributions}

\author{J\"orn Dunkel, Peter Talkner and Peter H\"anggi}

\address{Institut f\"ur Physik, Universit\"at Augsburg,
 Theoretische Physik I,  Universit\"atsstra{\ss}e~1, D-86135 Augsburg,
 Germany}
\ead{joern.dunkel@physik.uni-augsburg.de}

\begin{abstract}
The thermodynamic maximum principle for the Boltzmann-Gibbs-Shannon (BGS) entropy is reconsidered by combining 
elements from group and measure theory. Our analysis starts by noting that 
the BGS entropy is a special case of  
relative entropy. The latter characterizes probability distributions 
with respect to a pre-specified reference measure. To identify the canonical 
BGS entropy with a relative entropy is appealing for two reasons: 
(i) the maximum entropy principle assumes a coordinate invariant form; (ii) thermodynamic equilibrium distributions, which are obtained as solutions of the maximum entropy problem, may be characterized in terms of the transformation properties of the underlying reference measure (e.g., invariance under group 
transformations). As examples, we analyze two frequently considered candidates 
for the one-particle equilibrium velocity distribution of an ideal gas of relativistic particles. It becomes evident that the standard J\"uttner distribution is related to the (additive) translation group on momentum space. 
Alternatively, imposing Lorentz invariance of the reference measure leads to a so-called modified J\"uttner function, which differs from the standard 
J\"uttner distribution by a prefactor, proportional to the inverse particle energy.

\end{abstract}

\pacs{02.50.Cw, 05.40.-a, 05.70.-a}
%\submitto{\NJP}
\maketitle

%%%%%%%%%%%%%%%%%%%%%%%%%%%%%%%%%%%%%%%%%%%%%%%%%%
\section{Introduction}
%%%%%%%%%%%%%%%%%%%%%%%%%%%%%%%%%%%%%%%%%%%%%%%%%%

The combination of variational principles and group symmetries has proven extremely fruitful in various fields of theoretical physics over the past century, with applications ranging from classical mechanics~\cite{Goldstein,1993De} to quantum field theory~\cite{WeinbergQFT1,Coleman,Quigg}. In this paper, we would like to discuss how group and measure theoretical concepts may be incorporated into the maximum entropy principle (MEP) of canonical equilibrium thermostatistics~\cite{1957Ja}.\footnote{The MEP states that the thermodynamic equilibrium distribution can be obtained by maximizing a suitably chosen entropy functional under a given set of constraints.} To this end, we follow up an idea by Ochs~\cite{1976Ochs,1976Ochs_2} who demonstrated that the canonical Boltzmann-Shannon-Gibbs (BGS) entropy is a special case of \emph{relative entropy} (Sec.~\ref{section2}). The relative entropy~\cite{1978We} characterizes a probability distribution with respect to a prespecified reference measure and allows a manifestly coordinate invariant formulation of the MEP. In particular, we will focus on how the choice of the reference measure affects the solution of the entropy maximization problem (i.e., the equilibrium distribution). Thereby, it will be clarified that an acceptable MEP must include a postulate that determines which specific reference measure has to be used for a given class of physical systems. To obtain a mathematically meaningful characterization of potential reference measures, one can study their symmetry properties by means of their transformation behavior under group actions. The idea of combining measure and group theory goes back to the Hungarian mathematician Alfred Haar~\cite{Ha33}. In the second part of the paper, this approach will be pursued in order to analyze the MEPs for two of the most frequently discussed candidates for the relativistic one-particle equilibrium velocity distribution (Sec.~\ref{relativistic}).

%%%%%%%%%%%%%%%%%%%%%%%%%%%%%%%%%%%%%%%%%%%%%%%%%%
\section{Thermodynamic entropy, relative entropy and Haar measures}
%%%%%%%%%%%%%%%%%%%%%%%%%%%%%%%%%%%%%%%%%%%%%%%%%%
\label{section2}
We start out by summarizing the standard formulation of the canonical MEP in Sec.~\ref{standard}. We shall focus on the simplest paradigm, corresponding to a spatially homogeneous, ideal gases of classical particles, as this is sufficient for illustrating the main ideas. The concept of relative entropy is reviewed in Sec.~\ref{Relative_entropy}. The choice of the reference measures and their characterization in terms of symmetry groups is discussed in Sec.~\ref{Haar_measures}.

%%%%%%%%%%%%%%%%%%%%%%%%%%%%%%%%%%%%%%%%%%%%%%%%%%
\subsection{Standard formulation of the maximum entropy principle}
%%%%%%%%%%%%%%%%%%%%%%%%%%%%%%%%%%%%%%%%%%%%%%%%%%
\label{standard}
The canonical one-particle equilibrium velocity distribution for a non-relativistic gas of weakly interacting particles (e.g., atoms or molecules) is the Maxwell distribution, corresponding to the normalized probability density function (PDF)
\be\label{e:maxwell}
f_\mrm{M}(\bs v)=\left(\f{m\gb}{2\pi}\right)^{d/2} e^{-\gb  m\bs v^2/2},
\qquad\qquad
\bs v=(v_1,\ldots,v_d)\in \V_d:=\R^d.
\ee
Here, $\gb=1/T$ is the inverse temperature, $m$ the mass of the particle, and $\V_d$ denotes the space of the $d$-dimensional Cartesian velocity coordinates (throughout, we use units such that the speed of light $c=1$, and Boltzmann constant $\kB=1$). In principle, one can find several different arguments to justify Eq.~\eref{e:maxwell}~\cite{Becker}; e.g., it can be shown~\cite{1996MaKrSz} that the marginal one-particle PDF of an isolated, weakly interacting $N$-particle gas converges to $f_\mrm{M}(\bs v)$ in the thermodynamic limit. An alternative derivation that will be focussed on in the remainder of this paper is based on the canonical maximum entropy principle~(MEP). 
\par
The  MEP approach starts from postulating a canonical Boltzmann-Gibbs-Shannon entropy functional of the form
\numparts
\be\label{e:max-boltzmann_entropy}
S_\mrm{B}[f]:=-\int_{\V_d} \diff^d\bs v\,f(\bs v)\,\log [f(\bs v)\cdot c_0^d],
\ee
where $\diff^d\bs v=\diff v_1\ldots \diff v_d$ denotes an infinitesimal volume element of the non-relativistic Cartesian velocity space $\V_d$. The velocity constant $c_0$ is formally required in Eq.~\eref{e:max-boltzmann_entropy} to make the argument of the logarithm dimensionless; however, its actual value is irrelevant for the results to obtained, so that without loss of generality we can fix $c_0=c=1$ throughout. 
The MEP associated with Eq.~\eref{e:max-boltzmann_entropy} states that the thermodynamic equilibrium distribution $f^*$ is obtained by maximizing $S_\mrm{B}[f]$ with respect to $f\ge 0$ under the normalization and mean energy constraints
\be\label{e:max-constraints}
1=\int_{\V_d} \diff^d\bs v\,f(\bs v),\qquad \qquad 
\eps=\int_{\V_d} \diff^d\bs v\,f(\bs v)\;E(\bs v).
\ee
\endnumparts
Here, $E(\bs v)=m\bs v^2/2$ is the non-relativistic kinetic energy of a single 
particle (measured in the lab-frame), and $\eps$ the mean energy per particle which is assumed to be known. By means of two Lagrangian multipliers $(\ga,\gb)$, the MEP results in the condition
\be\label{e:max-lagrangian}
\nonumber
\fl\qquad
0&\equiv 
\f{\gd }{\gd f}\left\{
S_\mrm{B}[f]+
\ga\left[1 - \int_{\V_d} \diff^d\bs v\,f(\bs v)\right]
+\gb \left[\eps - \int_{\V_d} \diff^d\bs v\,f(\bs v)\;E(\bs v)\right]
\right\}\biggl|_{f=f^*}\\
\fl\qquad
 &= -(1+\log f^*)-\ga -\gb E.
\ee
Solving this equation for $f^*$ and determining $(\ga, \gb)$ from the constraints~\eref{e:max-constraints}, one recovers the Maxwellian~\eref{e:maxwell} with parameter $\gb=d/(2\eps)$. Hence, the MEP based on Eq.~\eref{e:max-boltzmann_entropy} appears to be satisfactory at first sight, but a more careful analysis reveals the following drawback: 
In order to give the empirically established result~\eref{e:maxwell}, the BGS entropy~\eref{e:max-boltzmann_entropy} must be written in terms of the \lq correct\rq\space physical variable, and one has to use the \lq correct\rq\space coordinate representation (in the above case, $\bs v$, or some linear transformation as momentum $\bs p=m\bs v$, expressed in Cartesian coordinates). Otherwise, one does not obtain the correct one-particle equilibrium distribution~\eref{e:maxwell}.  
\par
To briefly illustrate this, consider the physically most relevant three-dimensional case ($d=3$) and suppose that, instead of Cartesian coordinates $(v_1,v_2,v_3)$, we had started from polar coordinates $(v,\phi,\gt)\in [0,\infty)\times [0,2 \pi)\times [0,\pi]=:\PP^3$, i.e., by naively writing 
\numparts
\be\label{e:max-boltzmann_entropy-polar}
S_\mrm{B}[\bar f]:=
-\int_{\PP^3} \diff v\,\diff \phi\,\diff \gt\; \bar f(v,\phi,\gt)\,
\log \bar f( v,\phi,\gt),
\ee
where $\bar f(v,\phi,\gt)$ is subject to the constraints
\be\label{e:constraints-polar}
\fl\qquad
1=\int_{\PP^3} \diff v\,\diff \phi\,\diff \gt\; 
\bar f(v,\phi,\gt),
\qquad \qquad 
\eps=\int_{\PP^3}\diff v\,\diff \phi\,\diff \gt\; 
\bar f(v,\phi,\gt)\;\bar E(v,\phi,\gt) ,
\ee
\endnumparts
and $\bar E(v,\phi,\gt)=mv^2/2$ is the energy expressed in polar coordinates. Maximizing $S_\mrm{B}[\bar f]$ under the constraints~\eref{e:constraints-polar} yields
\be\label{e:polar_solution}
\bar f^*(v,\phi,\gt)=\left(\f{m\gb}{2\pi}\right)^{d/2} e^{-\gb  m v^2/2}.
\ee 
For comparison, by transforming the Maxwell PDF~\eref{e:maxwell} to polar coordinates we find
\be\label{e:maxwell_polar_solution}
\bar f_\mrm{M}(v,\phi,\gt)= 
\bar J\cdot \left(\f{m\gb}{2\pi}\right)^{d/2} e^{-\gb  m v^2/2}, 
\qquad\qquad
\bar J=v^2 \sin\gt, 
\ee
where $\bar{J}$ is the Jacobian of the coordinate transformation $(v_1,v_2,v_3)\mapsto (v,\phi,\gt)$. Upon comparing Eqs.~\eref{e:maxwell_polar_solution} and~\ref{e:polar_solution}, we observe that $\bar f^*\ne \bar f_\mrm{M}$, due to the missing Jacobian pre-factor in Eq.~\eref{e:polar_solution},
\par
This simple example illustrates that the above entropy definition is implicitly coordinate dependent. This fact is somewhat unsatisfactory. If viewed as fundamental, then the MEP should be formulated in a form that works independently from the underlying coordinate representation. As we shall discuss next, this can be achieved by recognizing that the thermodynamic entropy~\eref{e:max-boltzmann_entropy} is a special case of the so-called \emph{relative entropy}~\cite{1976Ochs,1976Ochs_2,1978We}.

%%%%%%%%%%%%%%%%%%%%%%%%%%%%%%%%%%%%%%%%%%%%%%%%%%
\subsection{Relative entropy} 
\label{Relative_entropy}
%%%%%%%%%%%%%%%%%%%%%%%%%%%%%%%%%%%%%%%%%%%%%%%%%%
First, we summarize the definition of the relative entropy~\cite{1976Ochs,1976Ochs_2,1978We} and demonstrate its invariance under coordinate transformations. Subsequently, it will be shown how the BGS entropy~\eref{e:max-boltzmann_entropy} is embedded into this concept. 
\par
Consider some set \mbox{$\X\subseteq \R^d$} and two measures $\mu$  and $\nu$ on $\X$ that are absolutely continuous with respect to each other (i.e., $\mu$ and $\nu$ have the same null sets in $\X$~\cite{Halmos51}). The \emph{relative entropy} of $\mu$ with respect to $\nu$ is defined by\footnote{This definition of \emph{relative entropy}, sometimes also referred to as generalized Boltzmann-Shannon-Gibbs entropy~\cite{1976Ochs,1976Ochs_2,1978We}, does not only apply to continuous but also to discrete spaces $\X$. Depending on the problem under consideration, $\X$ may be the one-particle or many-particle phase space, the velocity space, the configuration space, etc..}
\numparts
\be \label{e:relative_entropy_0}
S[\mu|\nu] 
:=
-\int_{\X} \diff \mu\,\log\f{\diff \mu}{\diff \nu}(\bs x)
=
-\int_{\X} \diff \nu\,f_{\mu|\nu}(\bs x)\,\log f_{\mu|\nu}(\bs x),
\ee
where the function
\be
f_{\mu|\nu}(\bs x):=\f{\diff \mu}{\diff \nu}(\bs x)>0
\ee
\endnumparts
is the so-called Radon-Nikodym density~\cite{Halmos51} of $\mu$ with respect 
to $\nu$. The measure $\nu$ plays the role of a \emph{reference measure}. 
We briefly illustrate the meaning of the Radon-Nikodym density by two simple examples:
\par
The most prominent measure on $\R^d$ is the Lebesgue measure, denoted by $\gl$~\cite{Halmos51}. The measure $\gl$ assigns to any $d$-dimensional rectangular parallel-epiped $\mcal{I}_d:=[a_1,b_1]\times\ldots\times [a_d,b_d]\subset \R^d$ the intuitive measure number $\gl(\mcal{I}_d)=\prod_{i=1}^d (b_i-a_i)$, where it is assumed that $b_i>a_i$ holds $\fa i=1,\ldots,d$. 
If, for example, $\mu$ is a probability measure on $\X\subseteq\R^d$, then the Radon-Nikodym density $f_{\mu|\gl}(\bs x)$ of $\mu$ with respect to $\gl$ is the \lq ordinary\rq\space PDF of $\mu$. 
\par
As the second example, consider two measures $\mu,\nu$ on $\X\subseteq\R^d$ with non-vanishing densities $f_{\mu|\gl}> 0$ and $f_{\nu|\gl}> 0$ on $\X$. In this case, the Radon-Nikodym density of $\mu$ with respect to $\nu$ is given by the quotient of their densities, i.e.,
\be\label{e:radon_nikodym}
f_{\mu|\nu}(\bs x):=
\f{\diff \mu}{\diff \nu}(\bs x)
=
\f{f_{\mu|\gl}(\bs x)}{f_{\nu|\gl}(\bs x)}.
\ee
Accordingly, we may rewrite the relative entropy~\eref{e:relative_entropy_0} in terms of the two densities  $f_{\mu|\gl}$ and $f_{\nu|\gl}$ as
\be\label{e:relative_entropy}
S[\mu|\nu] 
&=\nonumber 
-\int_{\X} \diff\gl\; 
f _{\mu|\gl}(\bs x)\, \log \f{f_{\mu|\gl}(\bs x)}{f_{\nu|\gl}(\bs x)}\\
&=
-\int_{\X} \diff^d\bs x\; 
f _{\mu|\gl}(\bs x)\, \log \f{f_{\mu|\gl}(\bs x)}{f_{\nu|\gl}(\bs x)}
=:s[f _{\mu|\gl}|f _{\nu|\gl}].
\ee
In the second line, we have inserted the equivalent notation $\diff^d\bs x$ for the Lebesgue measure $\diff \gl$ of an infinitesimal volume element in $\R^d$. Equation~\eref{e:relative_entropy} will provide the basis for all subsequent considerations. We note that, in order to define relative entropy, it is  \textit{a~priori} not required that the measures  $\mu$ and $\nu$ are normalizable on $\X\subseteq\R^d$; it suffices to assume that they have the same null sets, i.e., $f_{\mu|\gl}(\bs x)=0$ implies $f_{\nu|\gl}(\bs x)=0$ and vice versa, so that the argument of the logarithm is well-defined.\footnote{Conventionally, we set $\f{\diff \mu}{\diff \nu}(\bs x)=1$, if $f_{\mu|\gl}(\bs x)=f_{\nu|\gl}(\bs x)=0$.} 
\par
Before discussing how the BGS entropy~\eref{e:max-boltzmann_entropy} arises as a special case of  Eq.~\eref{e:relative_entropy}, it is useful to give the general, coordinate invariant form of the MEP with Eq.~\eref{e:relative_entropy} serving as the starting point. For this purpose, we impose the constraints
\numparts
\be\label{e:norm_constraint}
1&=\int_{\X} \diff \mu
&=\int_\X\diff^d\bs x\; f_{\mu|\gl}(\bs x),\\
\eps\label{e:energy_constraint}
&=\int_{\X} \diff \mu\;E(\bs x)
&=\int_\X\diff^d\bs x\; f_{\mu|\gl}(\bs x)\;E(\bs x),
\ee
\endnumparts
where $E>0$ is a non-negative \lq energy\rq\space function.\footnote{In principle, one could also include more than two constraints.} Maximizing $S[\mu|\nu]=s[f _{\mu|\gl}|f _{\nu|\gl}]$ with respect to $\mu$ or, equivalently, with respect to $f_{\mu|\gl}$, and taking into account  the constraints \eref{e:norm_constraint} and \eref{e:energy_constraint}, leads to the condition
\be\label{e:variation}
0\equiv 1+\log \f{f_{\mu|\gl}^*}{f_{\nu|\gl}} +\ga +\gb E,
\ee
Similar to Eq.~\eref{e:max-lagrangian}, $\ga$ and $\gb$ have entered here as Lagrangian multipliers for the normalization and \lq energy\rq\space constraints, respectively. From Eq.~\eref{e:variation} the solution of the variational problem is obtained as
\be\label{e:solution}
f_{\mu|\gl}^*(\bs x)=f_{\nu|\gl}(\bs x)\;e^{-(\ga+1) -\gb E(\bs x)}.
\ee
The parameters $(\ga,\gb)$ are determined by means of the conditions~\eref{e:norm_constraint} and \eref{e:energy_constraint}. As it is evident from Eq.~\eref{e:solution}, the \lq equilibrium\rq\space PDF $f_{\mu|\gl}^*$ depends on the choice of the reference density $f_{\nu|\gl}(\bs x)$. 
\par 
We next show that the relative entropy definition~\eref{e:relative_entropy} is manifestly coordinate invariant. For this purpose, consider a change of coordinates $\bs x\mapsto \bar{\bs x}$, and denote by $\bar{\X}$ the range of the new coordinates. Using the following standard formulae for the transformation of volume elements and densities $f$:
\be
\diff^d\bs x=\left(\f{\p\bs x}{\p\bar{\bs x}}\right)\;\diff^d \bar{\bs x},  
\qquad\qquad
\bar{f}(\bar{\bs x})=\left(\f{\p\bs x}{\p\bar{\bs x}}\right)
 f(\bs x(\bar{\bs x})),
\ee
where $\bar{J}=({\p\bs x}/{\p\bar{\bs x}})$ is the Jacobian of the coordinate transformation, we find that
\be\label{e:relative_entropy_transformed}
s[f _{\mu|\gl}|f _{\nu|\gl}]
&= \nonumber
-\int_{\X} \diff^d\bs x\; 
f _{\mu|\gl}(\bs x)\, \log \f{f_{\mu|\gl}(\bs x)}{f_{\nu|\gl}(\bs x)}\\
&= 
-\int_{\bar{\X}} \diff^d\bar{\bs x}\; 
\bar f _{\mu|\gl}(\bar{\bs x})\, 
\log \f{\bar f_{\mu|\gl}(\bar{\bs x})}{\bar f_{\nu|\gl}(\bar{\bs x})}
=s[\bar f _{\mu|\gl}|\bar f _{\nu|\gl}].
\ee
Hence, the relative entropy is indeed independent of the choice of the coordinates, due the fact that the Jacobians in the argument of the logarithm cancel. As a consequence, the solution of the associated MEP becomes coordinate independent as well. To demonstrate this more explicitly, we first rewrite the constraint function $E$ in terms of the new coordinates by defining $\bar{E}(\bar{\bs x}):=\bar{E}(\bs x(\bar{\bs x}))$. Then, the constraints~\eref{e:norm_constraint} and~\eref{e:energy_constraint} may be expressed equivalently in the new coordinates as
\numparts
\be\label{e:norm_constraint_new}
1
&=\int_\X\diff^d\bs x\; f_{\mu|\gl}(\bs x)
&=\int_{\bar{\X}}\diff^d\bar{\bs x}\; \bar f_{\mu|\gl}(\bar{\bs x}),\\
\eps\label{e:energy_constraint_new}
&=\int_\X\diff^d\bs x\; f_{\mu|\gl}(\bs x)\;E(\bs x)
&=\int_{\bar{\X}}\diff^d\bar{\bs x}\; \bar f_{\mu|\gl}(\bar{\bs x}) \;
\bar{E}(\bar{\bs x}).
\ee
\endnumparts
Hence, the solution of the associated variational problem reads
\be
\bar f_{\mu|\gl}^*(\bar{\bs x})
&=\nonumber
\bar{f}_{\nu|\gl}(\bar{\bs x})\;
e^{-(\ga+1) -\gb \bar{E}(\bar{\bs x})}\\
&=\nonumber
\left(\f{\p\bs x}{\p\bar{\bs x}}\right)
{f}_{\nu|\gl}(\bs x(\bar{\bs x}))\;
e^{-(\ga+1)-\gb E(\bs x(\bar{\bs x}))} \\
&=
\left(\f{\p\bs x}{\p\bar{\bs x}}\right)
{f}_{\mu|\gl}^*(\bs x(\bar{\bs x})).
\ee
This is indeed the correct transformation law for the equilibrium PDF ${f}_{\mu|\gl}^*$ from Eq.~\eref{e:solution}; 
i.e., once the reference measure $\nu$ and its density are properly specified, the MEP and its solution become independent of the choice of the coordinates.
\par
Finally, it is straightforward to see that the BGS entropy~\eref{e:max-boltzmann_entropy} is a special case of Eq.~\eref{e:relative_entropy}: We identify $\X=\V_d=\R^d$ and \emph{fix the reference measure as the Lebesgue measure in velocity space} $\nu=\gl$. Then, taking into account that $f_{\gl|\gl}(\bs v)\equiv 1$, Eq.~\eref{e:relative_entropy} reduces to the BGS entropy~\eref{e:max-boltzmann_entropy}; 
i.e., explicitly,
\be\label{e:relative_entropy_nonrelativistic}
S[\mu|\gl]
=
-\int_{\V_d} \diff^d\bs v\;f_{\mu|\gl}(\bs v)\,\log f_{\mu|\gl}(\bs v)
= 
S_\mrm{B}[f].
\ee
We thus note that the canonical BGS entropy corresponds to a specific choice of the reference measure, namely, the Lebesgue measure in velocity space. Put differently, whenever one writes an entropy in the \lq standard\rq\space form~\eref{e:relative_entropy_nonrelativistic}, one has implicitly fixed an underlying reference measure (defined with respect to some set of primary variables).  
\par
With regard to the subsequent discussion it will be important to keep in mind that the solution~\eref{e:solution} of the coordinate invariant MEP is determined by two ingredients: (i) the \lq energy\rq\space function $E$ that specifies the mean value constraint; (ii) the underlying reference measure $\nu$. While usually the energy function $E$ is known, it is a not-so-trivial problem to identify the appropriate reference measure $\nu$ for a given class of physical systems. In the next section, we are going to discuss how one can classify reference measures according their transformation properties under symmetry groups.

%%%%%%%%%%%%%%%%%%%%%%%%%%%%%%%%%%%%%%%%%%%%%%%%%%
\subsection{Choice of the reference measure: Group invariance and Haar measures}
\label{Haar_measures}
%%%%%%%%%%%%%%%%%%%%%%%%%%%%%%%%%%%%%%%%%%%%%%%%%%
The above discussion shows that the MEP is incomplete unless one is able to specify the reference measure $\nu$ on the state space  $\X$. Put differently, before accepting the MEP as a truly fundamental principle, one has to find a general method that allows to determine $\nu$ for a given class of dynamical systems. A promising step towards solving this problem is to analyze potential reference measures with respect to their invariance properties under fundamental symmetry transformations. Conceptually, this idea is closely related to the theory of Haar measures~\cite{Ha33, Halmos51}. In a seminal paper~\cite{Ha33} published in 1933,  the Hungarian mathematician Alfred Haar studied the possibility to introduce a measure $\mu_\circ$ \emph{on} a continuous group $(\G,\circ)$ such that $\mu_\circ$ is invariant under the group multiplication~\lq $\circ$\rq. To briefly sketch this idea, consider a subset $\mcal{A}$ of the group $\G$ and some arbitrary, fixed group element $g\in \G$. By multiplying each element $a\in\mcal{A}$  with $g$, the subset $\mcal{A}$ is mapped onto another subset of $\G$, denoted by
\be
g\circ\mcal{A}:=\left\{g\circ a\;|a\in\mcal{A}\;\right\}.
\ee
Now consider a measure $\mu_\circ$ on $\G$ that assigns to $\mcal{A}\subseteq\G$ some non-negative real number $\mu_\circ(\mcal{A})$. The measure $\mu_\circ$ is said to be group invariant, if\footnote{In the case of non-commutative (i.e., non-Abelian) groups, one may distinguish invariance under multiplications from the right or left.}
\be
\mu_\circ(g\circ\mcal{A})=\mu_\circ(\mcal{A})
\ee
holds for any $g\in \G$ and $\mcal{A}\subseteq\G$. Haar was able to prove the existence of an invariant measure $\mu_\circ$, and its uniqueness apart from an irrelevant multiplicative constant for locally compact, topological groups. Such group invariant measures $\mu_\circ$ are referred to as Haar measures nowadays~\cite{Halmos51}. They give a mathematically precise meaning to the notion \lq uniform distribution\rq\space by combining measure and group theoretical concepts. However, in physics one often encounters the slightly different situation, where a certain symmetry group acts on the domain $\X$ of a vector space, e.g., by means of a matrix representation. In this case, it is a natural to extend the original ideas of Haar by considering measures on $\X$ that are invariant under the group action.\footnote{For example, in the one-dimensional case $d=1$  the proper-orthochronous Lorentz group~$\mcal{L}_+^\uparrow$ consists of boosts only and, therefore, it can be identified with the relativistic velocity space $\RV_1=(-1,1)$; hence, the action of $\mcal{L}_+^\uparrow$ on $\RV_1$ is just the action of $\mcal{L}_+^\uparrow$ on itself. This corresponds to the framework originally considered by Haar~\cite{Ha33}. By contrast, in higher space dimensions $d>1$ it is not possible anymore to identify the relativistic  velocity space $\RV_d=\left\{\bs v\in \R^d\;|\;|\bs v|<1\right\}$ directly with a subgroup of the Lorentz group, since then the number of group parameters is larger than $d$ (cf. Chap. 6 in Ref.~\cite{SexlUrbantke}). Nevertheless, also in this case one can find a Lorentz invariant measure on $\RV_d$, which is unique apart from an irrelevant multiplicative constant; cf. discussion in Sec.~\ref{relativistic}.}
\par
 In order to link these concepts to thermodynamics, we return to the BGS entropy~\eref{e:relative_entropy_nonrelativistic}. This \lq canonical\rq\space entropy was identified above as the relative entropy with respect to the Lebesgue measure $\gl$ on the non-relativistic velocity space $\V_d=\R^d$. Adopting the group-theoretical point of view, the \emph{defining} property of the Lebesgue measure is given by the fact that  $\gl$ is the only\footnote{We omit the phrase \lq apart from an irrelevant multiplicative constant\rq\space from now on.} translation invariant measure on $\V_d$. To capture this fact more precisely, we define $\bs w$-parameterized translations $G_{\bs w}$ on $\V_d$ by means of
\be\label{e:Galilei}
G_{\bs w}:\bs v\mapsto \bs w+\bs v,
\qquad\qquad 
\fa \bs w,\bs v\in \V_d.
\ee 
The velocity translations $G_{\bs w}$ form a group by means of the composition rule
\be
G_{\bs w_1}\circ G_{\bs w_2}=G_{\bs w_1+\bs w_2}.
\ee
Now consider some subdomain $\mcal{A}\subset\V_d$ and define the translation $G_{\bs w}[\mcal{A}]$ of $\mcal{A}$ by
\be\label{e:Galilei_translation}
G_{\bs w}[\mcal{A}]:=\left\{ \bs w+\bs v\;|\;\bs v\in \mcal{A}\right\}.
\ee
Then the Lebesgue measure $\gl$ is the only measure satisfying~\cite{Halmos51}
\numparts
\be
\gl(G_{\bs w}[\mcal{A}])=\gl(\mcal{A}),
\qquad\qquad 
\forall
\mcal{A}\subset\V_d,\;\bs w\in\V_d,
\ee
or, equivalently, in differential notation
\be
\diff^d\bs v=\diff^d\bs v'
\ee
\endnumparts
where $\bs v'=G_{\bs w}(\bs v)=\bs w+\bs v$. This very translation invariance distinguishes the Lebesgue measure from all other measures that can be introduced on $\V_d$. Adopting the physicist's point of view, the translation map~\eref{e:Galilei} may be interpreted in two different ways.
\begin{enumerate}
\item \textbf{Geometric interpretation:} Equation~\eref{e:Galilei} describes a Galilei transformation, corresponding to a change of inertial coordinate systems. 
\item \textbf{Kinetic interpretation:} Equation~\eref{e:Galilei} describes a momentum conservation law, with $\Delta\bs p=m\bs w$ corresponding to the particle's momentum gain in a collision.
\end{enumerate}
Both interpretations are equally plausible here, because non-relativistic momentum and velocity differ by a mass constant $m$ only; in particular, the Lebesgue measure in velocity space transforms to a Lebesgue measure in momentum space, when changing from  velocity to momentum coordinates in the non-relativistic case. However, regardless of this ambiguity in the interpretation of Eq.~\eref{e:Galilei}, it is evident that the Lebesgue measure in velocity space (or, equivalently, in momentum space) plays a distinguished role in non-relativistic physics: \emph{It is the Haar measure of the Galilei group (or, equivalently, of the momentum translation group)}. This might explain why only the relative entropy with respect to this particular measure, $S[\mu|\gl]$, yields the correct non-relativistic equilibrium distribution~\eref{e:maxwell}. 
\par
In the remainder of this paper, we are going to study generalizations of the Maxwell distribution~\eref{e:maxwell} in the framework of special relativity. In particular, we shall identify the reference measures underlying two of the most commonly considered relativistic one-particle equilibrium distributions.

%%%%%%%%%%%%%%%%%%%%%%%%%%%%%%%%%%%%%%%%%%%%%%%%%%
\section{Relativistic velocity distributions}
%%%%%%%%%%%%%%%%%%%%%%%%%%%%%%%%%%%%%%%%%%%%%%%%%%
\label{relativistic}

Six years after Einstein~\cite{Ei05a,Ei05b} had formulated his theory of special relativity, Ferencz J\"uttner~\cite{1911Ju}
presented in 1911 the first detailed study on the canonical thermostatistics
of a relativistic (quasi-)ideal gas of classical particles.\footnote{J\"uttner's paper~\cite{1911Ju} appeared about 20 years prior to Haar's work on group invariant measures~\cite{Ha33}.}  As the main result of his paper, he proposed the following three-dimensional relativistic generalization of
Maxwell's non-relativistic momentum distribution~\cite{1911Ju,1921Pa,1957Sy}:
\be\label{e:juettner}\label{e:juettner-a}
\phi_\mrm{J}(\bs p)=\f{\exp(-\gb E)}{\mcal{Z}_0},
\qquad\qquad
\bs p\in\RP_3:=\R^3
\ee
with $\gb=1/T$ being the inverse temperature parameter, and
\be
E=(m^2+\bs p^2)^{1/2}=m\gc(\bs v),
\qquad\qquad
\bs p= m\bs v\gc(\bs v)
\ee
the relativistic energy and the relativistic momentum with Lorentz factor $\gc(\bs v)=(1-\bs v^2)^{-1/2}$ (we continue to use units $\kB=c=1$). The $d$-dimensional relativistic momentum space is denoted by $\RP_d$. The constant $\mcal{Z}_0$ is determined by the normalization condition
\be
1=\int_{\R^3}\diff^3\bs p\;\phi_\mrm{J}(\bs p),
\ee
and, in the three-dimensional case, one finds~\cite{1911Ju}
\be
\mcal{Z}_0=4\pi m^3\,\f{K_2(\gb m)}{\gb m},
\ee
where $K_\nu$ denotes the modified Bessel function of the second kind. The one-particle momentum distribution~\eref{e:juettner-a} refers to a 
laboratory rest frame, where the container enclosing the gas is at rest. As usual, it is assumed that for an ordinary hard box potential the spatial part of the one-particle \emph{phase space} PDF is trivial (i.e., constant), corresponding to a spatially homogeneous particle distribution in the box. 
\par
\begin{figure}
\includegraphics[angle=0]{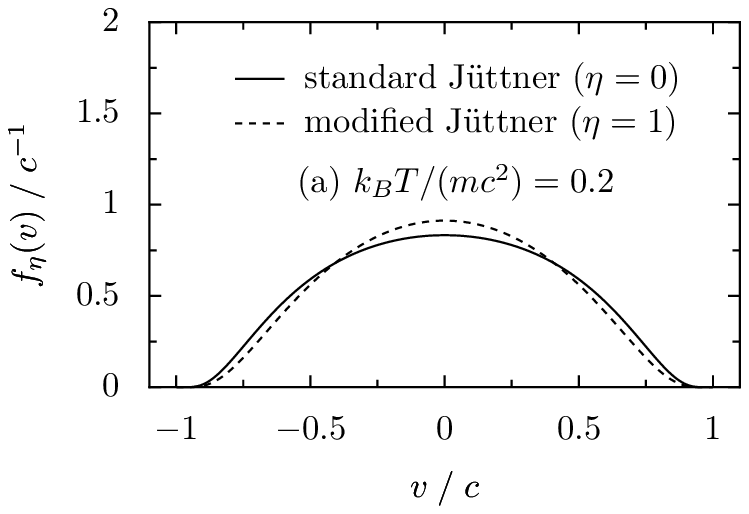}
\includegraphics[angle=0]{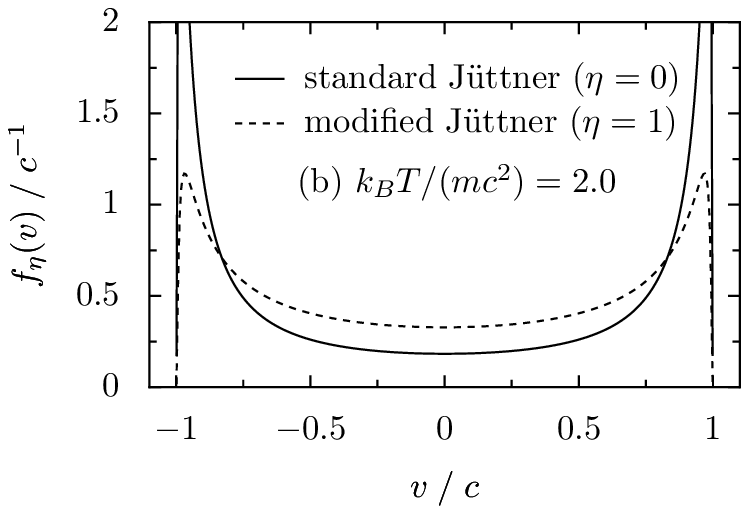}
\caption{\label{fig01}
Relativistic velocity PDFs from Eq.~\eref{e:mod_juettner_vel} for the one-dimensional case $d=1$. (a) At low temperatures $T\ll m$ both distributions approach a quasi-Gaussian (Maxwellian) shape. (b) At high temperatures  $T\gg m$ the distributions become bimodal. For the same temperature value $T$, the standard J\"uttner distribution with $\eta=0$ exhibits a higher probability of large absolute velocities than the modified distribution with $\eta=1$. 
}
\end{figure}

The J\"uttner function~$\phi_\mrm{J}$ has been widely used in high energy and astrophysics over the past decades~\cite{1971TeWe,CercignaniKremer,Liboff}. However, in recent years several authors~\cite{1989HoShSc,1981HoScPi,2005Sc,2006Le,2007DuHa} argued that Eq.~\eref{e:juettner} might not represent the correct relativistic equilibrium distribution, and several alternatives were suggested. Generalizing to an arbitrary number of space dimensions~$d$, the proposed candidates can be summarized in terms of the following $\eta$-parameterized momentum PDF:
\be\label{e:mod-juettner}
\phi_\eta(\bs p)=\f{\exp(-\gb E)}{\mcal{Z}\,E^\eta},
\qquad\qquad
\bs p\in\RP_d=\R^d,
\qquad
\eta\ge 0.
\ee
The normalization constant $\mcal{Z}$ depends on both $\eta$ and $d$. For \mbox{$\eta=0$} the PDF~\eref{e:mod-juettner}  reduces to the standard J\"uttner function~\eref{e:juettner}, $\phi_0\equiv\phi_\mrm{J}$. The most frequently considered modification corresponds to $\eta=1$~\cite{1989HoShSc,1981HoScPi,2005Sc,2006Le,2007DuHa,2007DuTaHa}, while one author~\cite{2006Le} has also included the case $\eta=2$. Compared with the J\"uttner value $\eta=0$, larger values $\eta>0$ diminish the probability of particles having high absolute momentum at same temperature $T=1/\gb$. The one-particle velocity PDF corresponding to Eq.~\eref{e:mod-juettner} is given by 
\be\label{e:mod_juettner_vel}
f_\eta(\bs v)=\f{m^{d-\eta}}{\mcal{Z}}\,
\f{\gc^{2+d}(\bs v)}{\gc^{\eta}(\bs v)}\,\exp[-\gb m \gc(\bs v)]
\ee
with $\bs v$ taking values in the relativistic velocity space $\RV_d:=\left\{\bs v\in \R^d\;|\;|\bs v|<1\right\}$. Below we focus on the two most frequently considered values $\eta=0$ (standard J\"uttner distribution) and  $\eta=1$ (modified J\"uttner distribution). Figure \ref{fig01} shows the corresponding velocity PDFs $f_0$ and $f_1$ at two different temperature values for the one-dimensional case $d=1$.  
\par
In the remainder, we will analyze the MEPs that give rise to the standard and modified J\"uttner distributions, respectively. In particular, the different underlying reference measures shall be characterized by means of their invariance under group actions.
%%%%%%%%%%%%%%%%%%%%%%%%%%%%%%

%%%%%%%%%%%%%%%%%%%%%%%%%%%%%%%%%%%%%%%%%%%%%%%%%%
\subsection{Standard J\"uttner distribution: Momentum translation symmetry}
\label{Juettner}
%%%%%%%%%%%%%%%%%%%%%%%%%%%%%%%%%%%%%%%%%%%%%%%%%%

We first consider the MEP for the standard J\"uttner distribution with $\eta=0$.  As discussed in Sec.~\ref{Relative_entropy}, the MEP becomes coordinate independent if expressed in terms of relative entropy. In the relativistic case, it is most convenient to use the momentum coordinate $\bs p\in\RP_d:=\R^d$.  The Lebesgue measure on relativistic momentum space $\RP_d$ has, by definition, a constant density denoted by $\ell$. Without loss of generality, we choose the normalization $\ell(\bs p)=(mc)^{-d}=m^{-d}$ so that the integral of $\ell$ over some finite subset of $\RP_d$ is a dimensionless number. With these preliminaries, we can state the MEP for the standard J\"uttner function: Maximization of the relative entropy  
\numparts
\be 
s_0[\phi|\ell] 
%&=\nonumber
%-\int_\RP \diff^d \bs p\, \phi(\bs p) \log\biggl[\f{\phi(\bs p)}{\ell(\bs p)}
%\biggr]
=
-\int_{\RP_d} \diff^d \bs p\; \phi(\bs p) \log\biggl[m^d\phi(\bs p)\biggr]
\label{e:relative_entropy_relativistic} 
\ee 
under the constraints
\be\label{e:max-constraints_juettner}
1=\int_{\RP_d} \diff^d\bs p\;\phi(\bs p),\qquad \qquad 
\eps=\int_{\RP_d} \diff^d\bs p\;\phi(\bs p)\;E(\bs p),
\ee
\endnumparts
where now $E=(m^2+\bs p^2)^{1/2}$ is the relativistic energy, yields the standard J\"uttner distribution $\phi_\mrm{J}$, corresponding to $\eta=0$ in Eq.~\eref{e:mod-juettner}. It may be worth noting that, in the relativistic case, the Lebesgue measure on $\RP_d$ does \emph{not} transform into a Lebesgue measure on the relativistic velocity space $\RV_d$ due to the nonlinear momentum-velocity relation $\bs p=m\bs v\gc(\bs v)$. Hence, if one rewrites the relative entropy $s_0$ in terms of the velocity $\bs v$, an additional determinant factor enters in the argument of the logarithm.
\par
We now turn to the invariance properties of the specific reference measure, required to obtain the standard J\"uttner distribution with $\eta=0$. Analogous to the discussion in Sec.~\ref{Haar_measures}, the Lebesgue measure in relativistic momentum space is singled out by the fact that it is the only translation invariant measure in momentum space; i.e., it is the Haar measure of the momentum translation group. Hence, the standard J\"uttner function is consistent with the \emph{kinetic interpretation} in Sec.~\ref{Haar_measures}. Put differently, if the J\"uttner function turns out to be the correct relativistic one-particle equilibrium distribution, then the maximum principle for the relative entropy  should be completed by the postulate that the reference measure must be translation invariant in momentum space.

%%%%%%%%%%%%%%%%%%%%%%%%%%%%%%%%%%%%%%%%%%%%%%%%%%
\subsection{Modified J\"uttner distribution: Lorentz symmetry}
%%%%%%%%%%%%%%%%%%%%%%%%%%%%%%%%%%%%%%%%%%%%%%%%%%

As the second example, we consider the modified J\"uttner distribution with $\eta=1$ in Eq.~\eref{e:mod-juettner}. It is straightforward to verify that this distribution is obtained by maximizing the relative entropy
\be 
s_1[\phi|\rho] 
=
-\int_{\RP_d} \diff^d \bs p\; \phi(\bs p) 
\log\biggl[\f{\phi(\bs p)}{\rho (\bs p)}\biggr],
\label{e:relative_entropy_relativistic_2}
\qquad\qquad
\rho(\bs p)=1/E 
\ee 
under the constraints~\eref{e:max-constraints_juettner}. In contrast to Eq.~\eref{e:relative_entropy_relativistic}, the reference density $\rho=1/E$ is momentum dependent. The measure $\chi$ associated with $\rho$ assigns to any subset 
$\mcal{A}\subset \RP_d$ the measure number
\be
\chi(\mcal{A})
=\int_\mcal{A}\diff^d \bs p\; \rho(\bs p)
=\int_\mcal{A}\f{\diff^d \bs p}{E(\bs p)}.
\ee 
It is interesting to explore the invariance properties of this measure. For this purpose, we consider an arbitrary proper-orthochronous Lorentz transformation. Such transformations are either spatial rotations, or boosts, or a combination of both~\cite{SexlUrbantke}. They act as linear transformations on the energy-momentum vector $(E,\bs p)$. Due to the fixed relation $E(\bs p)=(m^2+\bs p^2)^{1/2}$ between energy and momentum, a Lorentz transformation can also be viewed as transformation that operates on the momentum coordinates $\bs p$ alone, denoted by $L:\RP_d\to \RP_d$. The functions $L$ are linear only in the case of pure rotations, but nonlinear otherwise~\cite{1969VK}. However, analogous to Eq.~\eref{e:Galilei_translation}, we may define the Lorentz transformation $L[\mcal{A}]$ of a set $\mcal{A}\subset \RP_d$ by
\be
L[\mcal{A}]:=
\left\{ L(\bs p)\;|\;\bs p\in \mcal{A}\right\}.
\ee
By taking into account the well-known fact that~\cite{PeSc95,CercignaniKremer,1969VK}
\be\label{e:L-invariance}
\f{\diff^d \bs p}{E(\bs p)}=\f{\diff^d \bs p'}{E(\bs p')}
\ee
holds under Lorentz transformations, one then finds that
\be
\chi(L[\mcal{A}])=\chi(\mcal{A}),\qquad\qquad
\fa  \mcal{A}\in\RP_d.
\ee
Hence, the specific reference measure underlying the modified J\"uttner distributions with $\eta=1$ is distinguished by the property that it is Lorentz invariant. In view of the fact that the Lorentz group is the relativistic counterpart of the Galilei group, one can say that the modified J\"uttner distribution is obtained when adopting the~\emph{geometric interpretation} in Sec.~\ref{Haar_measures}. Put differently, if the modified J\"uttner function were the correct relativistic one-particle equilibrium distribution, then the maximum principle for the relative entropy  should be completed by the postulate that the reference measure  in momentum space must be invariant under the action of the fundamental symmetry group of the physical model (e.g., Galilei, Lorentz, etc.).

\paragraph{Explicit example: One-dimensional case $d=1$.}
As remarked earlier, the one-dimensional case $d=1$ is somewhat special, because (only) in this case the Lorentz boosts form a group that may be directly identified with the one-dimensional velocity space $\RV_1:=(-1,1)$. The composition of two Lorentz boosts induces a group multiplication $\oplus_\psi$ on
 $\RV_1$, given by
\be
v_3
:=
v_1\oplus_\psi v_2
:=\f{v_1+ v_2}{1+v_1 v_2}.
\ee
This group operation is well known as the Einstein addition of velocities. The task of introducing an invariant measure on the group $(\RV_1,\oplus_\psi)$ falls exactly into the class of problems originally considered by Haar~\cite{Ha33}. 
The subscript $\psi$ symbolizes that the Einstein addition $\oplus_\psi$ is equivalent to an ordinary addition \lq+ \rq \space in the space $\Psi:=(-\infty,\infty)$ of the rapidity variables \mbox{$\psi:=\atanh v$}. Put differently, the maps \lq$\atanh$\rq\space and \lq$\tanh$\rq\space induce a group isomorphism between  $(\RV_1,\oplus_\psi)$ and $(\Psi,+)$. The latter fact makes it particularly simple to identify the Haar measure on $(\RV_1,\oplus_\psi)$: One merely needs to rewrite the Lebesgue measure $\gl_\psi$ on $\Psi$, which is invariant under the addition of rapidities, in terms of the velocity coordinate; in differential notation, one then finds
\be\label{e:psi}
\diff \gl_\psi=\diff \psi
=
\left(\f{\diff \psi}{\diff v}\right) \diff v=\gc^2(v)\;\diff v
\propto \f{\diff p}{E(p)},
\ee
corresponding to the Lorentz invariant measure on $\RV_1$ and $\RP_1$, respectively  [cf. Eq.~\eref{e:L-invariance}]. As discussed above, using this measure as the reference measure in the MEP, one obtains the one-dimensional modified J\"uttner distribution with $\eta=1$.
\par 
For comparison, the ordinary addition $p_3:=p_1+p_2$ in momentum space $\RP_1=(-\infty,\infty)$ induces another group operation $\oplus_p$ on $\RV_1=(-1,1)$ by means of the map $v(p)=p/(m^2+p^2)^{1/2}$. The corresponding velocity addition law reads explicitly 
\be\label{juettner_law}
v_3
:=
v_1\oplus_p v_2
:=
\f{v_1\sqrt{1-v_2^2}+v_2\sqrt{1-v_1^2}}
{\sqrt{1-v_1^2 v_2^2+2v_1v_2\sqrt{1-v_1^2}\sqrt{1-v_2^2}}}.
\ee
Analogous to Eq.~\eref{e:psi}, the invariant Haar measure on $(\RV_1,\oplus_p)$ is obtained by expressing the Lebesgue measure $\gl_p$ on $\RP_1$, which is invariant under the momentum addition, in terms of the velocity variable, yielding
\be
\diff \gl_p=\diff p\propto \gc^3(v)\;\diff v.
\ee
As discussed in Sec.~\ref{Juettner}, by using this measure in the MEP one is led to the standard J\"uttner function.

\section{Summary} 
  
We have studied the canonical maximum entropy principle (MEP) for thermodynamic equilibrium distributions by combining basic ideas from group and measure theory~\cite{Ha33,Halmos51}. It has been demonstrated that the concept of relative entropy~\cite{1976Ochs,1976Ochs_2,1978We}  provides a suitable basis for stating the MEP in a coordinate invariant way. Moreover, this approach clarifies that thermodynamic equilibrium distributions, if obtained from a MEP~\cite{1957Ja}, are determined not only by their constraint functionals but also by the underlying reference measures. The latter may be characterized in terms of their symmetry properties, i.e., by their invariance under group actions. 
\par
As examples, we analyzed the two most frequently considered candidates~\cite{1911Ju,1921Pa,1957Sy,1971TeWe,CercignaniKremer,Liboff,1989HoShSc,1981HoScPi,2005Sc,2006Le,2007DuHa,2007DuTaHa} for the relativistic generalization of the Maxwell distribution. It could be shown that the two candidate distributions are based on different underlying reference measures. The reference measure leading to a standard J\"uttner distribution~\cite{1911Ju,1921Pa,1957Sy} is uniquely characterized by the fact that it is invariant under momentum translations, whereas the modified J\"uttner distribution~\cite{1989HoShSc,1981HoScPi,2005Sc,2006Le,2007DuHa,2007DuTaHa} is related to a Lorentz invariant reference measure in momentum space.
 Even though the above approach clarifies the underlying mathematical differences on a fundamental level, it does not permit to decide which distribution actually is the better candidate, as either reference measure has its own merits. In our opinion, this ambiguity deserves further consideration in the future. 
\par
We conclude this paper by mentioning two applications. 
The correct relativistic equilibrium
distribution is required in order to calculate the friction coefficients
and noise correlation functions of relativistic Langevin equations (RLEs)
self-consistently~\cite{2006DuHa}. An accurate determination of these
quantities is essential, e.g., if RLEs are employed to estimate the
outcome of high energy collision experiments, as recently done
by van Hees et al.~\cite{2006HeGrRa}. 
Another potential, astrophysical application concerns the Sunyaev-Zeldovich (SZ) effect \cite{1972SuZe,1998ItKoNo}, i.e., the distortion of the cosmic microwave background (CMB) spectrum due to the
interaction of CMB photons with the hot electron gas in
 clusters of galaxies. The size of relativistic corrections to the SZ effect depends essentially on the shape of the assumed relativistic electron velocity 
distribution.
\par
The authors would like to thank J.~Casado-Pascual, J.~Chluba, D.~Cubero, S.~Hilbert, K.~Sakmann, M.~Schindler, S. Weber, and M. Wubs for helpful discussions.

\section*{References}
\bibliographystyle{unsrt}
\bibliography{HaarMeasures}

\end{document}